\documentclass[%
aip,
sd,
amsmath,
amssymb,
reprint,
]{revtex4-2}
\allowdisplaybreaks
\usepackage{graphicx}
\usepackage{bm}
\usepackage{xspace}
\usepackage[utf8]{inputenc}
\usepackage[T1]{fontenc}
\usepackage{mathptmx}
\usepackage{etoolbox}
\usepackage{mathtools}

\makeatletter
\def\@email#1#2{%
 \endgroup
 \patchcmd{\titleblock@produce}
  {\frontmatter@RRAPformat}
  {\frontmatter@RRAPformat{\produce@RRAP{*#1\href{mailto:#2}{#2}}}\frontmatter@RRAPformat}
  {}{}
}%
\makeatother

\DeclareMathAlphabet{\mathcal}{OMS}{cmsy}{m}{n}
\SetMathAlphabet{\mathcal}{bold}{OMS}{cmsy}{b}{n}

\DeclareMathOperator{\Var}{Var}
\DeclareMathOperator{\Cov}{Cov}
\DeclareMathOperator{\E}{\mathbb{E}}
\newcommand{\e}{\mathrm{e}}
\renewcommand{\i}{\mathrm{i}}

\newcommand{\votwo}{VO$_2$\xspace} % VO2
\newcommand{\mJcm}{$\frac{\mathrm{mJ}}{\mathrm{cm}^2}$\xspace} % fluence unit
\newcommand{\Qb}{Q_\mathrm{B}} % bunch charge
\newcommand{\SNR}{\mathrm{SNR}} % signal-to-noise ratio

\newcommand{\ordSN}{(\Sigma\!\to\!N)}
\newcommand{\ordNS}{(N\!\to\!\Sigma)}

\newcommand{\Inorm}{I^\circ}

\begin{document}

\preprint{AIP/123-QED}

\title[Direct Electron Detectors in UED(S)]{Prospects for Direct Electron Detectors in Ultrafast Electron Diffraction and Scattering Experiments}

\author{Laurenz Kremeyer}
\affiliation{Department of Physics, Centre for the Physics of Materials, McGill University, Montreal, Québec H3A 2T8, Canada}
\email{laurenz.kremeyer@mail.mcgill.ca}
\author{David Cai}
\affiliation{Department of Physics, Centre for the Physics of Materials, McGill University, Montreal, Québec H3A 2T8, Canada}
\author{Malik Lahlou}
\affiliation{Department of Physics, Centre for the Physics of Materials, McGill University, Montreal, Québec H3A 2T8, Canada}
\author{Sebastian Hammer}
\affiliation{Department of Physics, Centre for the Physics of Materials, McGill University, Montreal, Québec H3A 2T8, Canada}
\affiliation{Department of Chemistry, McGill University, Montreal, Québec H3A 0B8, Canada}
\author{Raphael Schwenzer}
\affiliation{Fachbereich Physik, Universität Konstanz, Konstanz 78464, Germany}
\author{Bradley J. Siwick}
\affiliation{Department of Physics, Centre for the Physics of Materials, McGill University, Montreal, Québec H3A 2T8, Canada}
\affiliation{Department of Chemistry, McGill University, Montreal, Québec H3A 0B8, Canada}
\email{bradley.siwick@mcgill.ca}

\date{\today}

\begin{abstract}
Ultrafast electron diffraction and phonon-diffuse scattering [UED(S)] experiments make use of photo-induced changes to electron scattering intensity across 2D detectors to report on a very wide range of dynamic structural phenomena in molecules and materials.
Hybrid pixel counting detectors (HPCDs) are a promising technology for improved sensitivity and signal-to-noise in UED(S) experiments, as they offer near-zero readout noise and dark counts with the possibility of new acquisition modalities (e.g., shot-to-shot normalization) due to their high frame rates.
However, it is well known that HPCDs suffer from count losses at high electron fluxes even in CW beam applications.
How this translates to ultrashort electron pulse exposures has yet to be determined and is critical to understanding the application of this technology to ultrafast electron scattering experiments.
Here we show that count losses are significantly exacerbated in ultrafast (pulsed) experiments and that HPCDs require unconventional data handling and saturate above $\approx\!2$ electrons per pixel per pulse.
This count-rate limitation presents a severe constraint on electron bunch charge when interrogating single crystal samples.
Normalization strategies to optimize signal-to-noise in UED(S) and a complete model for measurement uncertainties using HPCDs are developed and tested using a large dataset.
Finally, we suggest ways HPCDs could be better adapted to ultrashort pulsed beam experiments.
\end{abstract}

\maketitle

\section{Introduction}
Ultrafast laser-pump, electron (or x-ray)-probe experiments have made it possible to directly measure structural and electronic dynamics in molecules and materials in the time-domain\cite{Filippetto2022Ultrafast,Sciaini2011Femtosecond} with sub-50\,fs resolution.
These nonequilibrium methods provide access to many of the microscopic processes that underlie material properties, including electron-phonon\cite{Maldonado2020Tracking,Tauchert2022Polarized,Britt2022Direct} and phonon-phonon coupling\cite{Harb2016Phonon,Kremeyer2024Ultrafast} as well as lattice, charge, and orbital ordering phenomena\cite{Ehrke2011Photoinduced,Cheng2024Ultrafast,Dornes2019Ultrafast}.
The ability to interrogate photo-induced phase transitions with these techniques has also deepened our understanding of how to optically prepare and control novel states of materials that exhibit properties that are inaccessible under equilibrium conditions\cite{Kogar2019Light, Morrison2014Photoinduced, Basov2017Towards}.
Compared to ultrafast spectroscopies, these techniques have very high structural-information content and competitive time resolution, but sensitivity to relative changes in electron scattering intensity is orders of magnitude lower.

\begin{figure*}[t]
\includegraphics[width=1\textwidth]{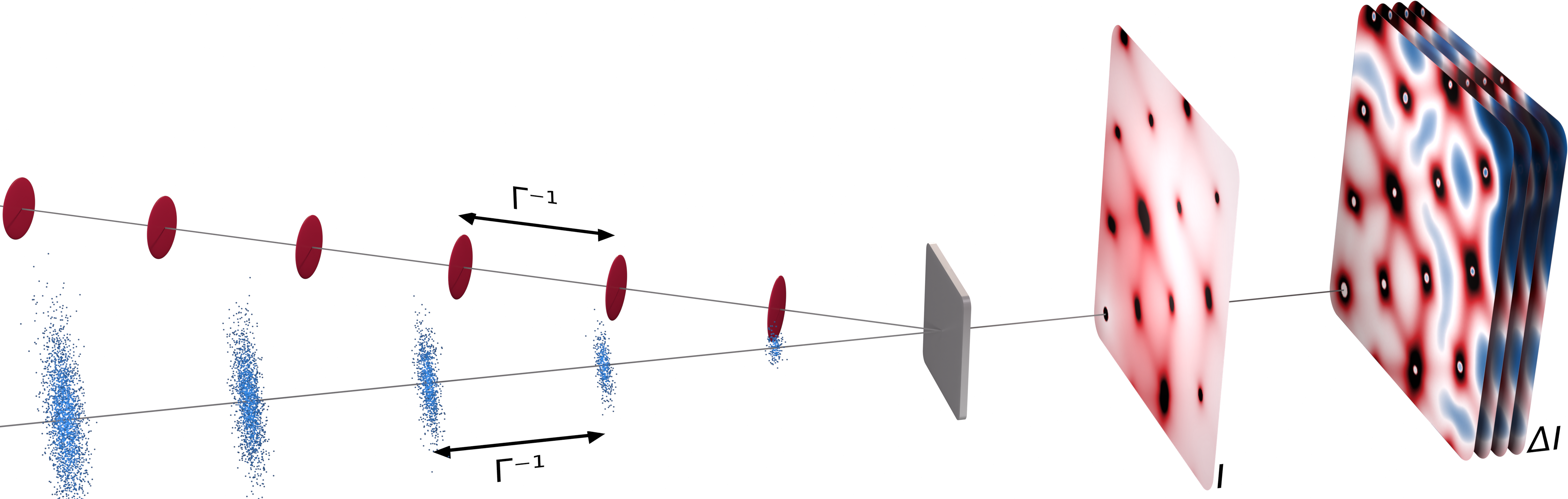}
\caption{Schematic illustration of a UED(S) experiment. The sample is first excited with an ultrashort pump laser pulse (red discs). The pump is followed by a probe electron pulse (blue particle clouds) at a well-controlled delay time, producing an electron intensity pattern on the detector (I). The repetition rate of electron bunches and optical pulses is typically determined by the laser amplifier and here $\Gamma^{-1}=1$\,ms. Images on the far right illustrate calculated difference images (pump-on minus pump-off, $\Delta$I) for simulated diffraction and phonon-diffuse scattering patterns of graphite at different delay times. Red (blue) regions represent increases (decreases) in electron intensity, the signals to be measured in UED(S) experiments.}
\label{fig:exp}
\end{figure*}

Here we focus on the potential of direct electron detectors to improve the sensitivity of ultrafast electron-probe experiments.
The default mode of data acquisition in these experiments is time-resolved scattering (Fig.~\ref{fig:exp}), with the scattering signals measured over a range of scattering-angles using a 2D detector.
Although time-resolved transmission electron imaging/microscopy is also under active development\cite{lagrange2025laser}, we will not discuss the particularities of time-resolved image resolution and sensitivity (detector quantum efficiency).
An essential feature of the raw scattering signals measured is that they can be distributed highly non-uniformly across the detector.
For gas-phase samples, polycrystalline materials or textured films, the dynamic range of the raw scattering signals tends to be narrow.
For example, in polycrystalline monoclinic \votwo, the dynamic range between the low vs high-index Debye-Scherrer ring intensities measured under typical conditions in an ultrafast electron diffractometer is only of the order $\approx\!10$.
When probing high-quality single crystal samples, however, scattering is concentrated into Bragg peaks and the dynamic range of the raw scattered intensity tends to be many orders of magnitude larger than that observed from polycrystalline, amorphous or gas-phase samples.
For example, under typical experimental conditions peak intensity in the first- and second-order Bragg reflections of graphite is $10^5$ times the intensity of the phonon-diffuse scattering features observable between the Bragg peaks\footnote{dynamic range extracted from data of\cite{RdC2019Time}} (Fig.~\ref{fig:exp}).

The relevant signals in UED(S) are the pump-probe delay-time dependent changes in electron scattering that result from photoexcitation of the sample, measured over as wide a range of scattering vector as possible.
Current state of the art measurements are capable of resolving differential scattering signals (i.e., pump-on minus pump-off) at the 1\%--10\% level relative to the unpumped scattering intensity, although signals approaching 0.1\% have been reported for high-intensity features (Bragg peaks)\cite{Cheng2022Light, Hammer2024Excimer}.
To achieve a signal-to-noise ratio (SNR) of 5 for a signal change of 1\%, differential scattering intensity measurements with a relative uncertainty (or relative standard error in the mean) of 0.002 are required.
If the measurement is shot-noise limited, this requires the feature in question to contain at least 250,000 counts.

It is worthwhile to compare these actually measured signals with those that can be expected under weak-excitation conditions in a system that does not undergo a structural phase transition.
For example, if laser-deposited energy results in a 10\,K increase in specimen temperature (after full thermalization) the (200), (300), and (400) Bragg peaks in graphite would be suppressed by 0.05\%, 0.16\%, and 0.25\%, respectively (at room temperature), while single phonon-diffuse scattering from low-frequency acoustic modes increases by approximately 3\% and optical modes by only 0.1\%--0.3\%.
Measurements of signal changes at the 0.1\% level (or even the 0.001\% level) are required to access the weak pump-excitation regime in UED(S) experiments.
Under shot-noise limits, every 10-fold improvement in the resolvable relative signal change (at constant SNR) requires a 100-fold increase in the number of counts contained in the measured feature.
Thus, a critical direction for improving the sensitivity of UED(S) measurements has been maximizing electron count rates and the primary argument favoring the development of high bunch charge and high repetition rate ultrafast electron sources\cite{Filippetto2022Ultrafast, domrose2025megahertz}.
However, stretching the dynamic range of detectors at both high and low count rates is also essential to enhance sensitivity, providing a strong argument in favor of improved detector technology.
The detector noise-floor, specifically, is the essential determining factor for low electron count rate UED(S) signals to be reliably measured (or measured at all).
HPCDs\cite{tate2016high,rodrigues2024high} operating in electron counting mode reduce readout and dark noise to near-zero which may open the possibility of shot-noise limited measurements even at the extremely low count rates associated with phonon-diffuse scattering signals in UED(S) measurements and other low scattering cross section inelastic processes in the nascent technique of ultrafast electron energy loss spectroscopy (uEELS)\cite{barantani2025ultrafast, kim2023using}.
Being true electron-counting detectors, HPCDs have a number of potential signal-to-noise advantages over other technologies in use (microchannel plate and phosphor scintillator-based detectors) that could lead to improved sensitivity in UED(S) experiments and significantly broaden the scope of this technique, particularly in the direction of weaker-excitation levels.

Here we report on the performance of HPCDs, specifically the Dectris Quadro, in ultrafast electron scattering experiments.
In this article we discuss the mechanisms and statistics of electron counting detectors for ultrafast electron diffraction in depth, revealing their limitations in ultrafast applications and presenting strategies to optimize SNRs using HPCDs.
Throughout, we express the electron dose received by the detector in terms of electrons/pulse per pixel (EPP), $\bar\lambda$.
In current instruments, the electron dose can range from many thousands in the main beam to tens in the brightest Bragg peaks of a diffraction pattern to arbitrarily small values in the diffuse features between Bragg peaks.
One can convert the electron dose from a per-pulse basis to a per-time basis by simply multiplying the electron dose in EPP by the laser repetition rate,~$\Gamma$.

The remainder of this paper is organized as follows.
In Sec.~\ref{sec:methods}, we describe the ultrafast electron source and detector used in this study.
In Sec.~\ref{sec:pixel_response}, we characterize the electron counting behavior of the detector at the individual pixel level when subject to ultrafast exposures at (average) count rates ranging from below 0.1 to greater than 10\,EPP in both detector modes of operation (normal and retrigger).
We also establish an estimator and associated uncertainty for the average count rate, which we call the $P_0$ counting method, that corrects for lost counts and extends the dynamic range of the detector at the upper end (saturation) by more than an order of magnitude compared to the conventional (direct) counting mode.

In Sec.~\ref{sec:source_noise}, we characterize the full experimental noise spectrum of our instrument while employing a HPCD and investigate the SNR performance of data (self) normalization strategies ranging from multi-pulse down to shot-to-shot. Here, we use a dataset that is not complicated by lost counts in conventional (direct) counting mode to focus solely on the SNR impact of data normalization independent of saturation/counting correction via the $P_0$ counting method. We demonstrate that shot-to-shot normalization, enabled by the high-frame readout rates of HPCDs, provides little benefit in terms of improved SNR at the levels of source noise present in typical UED(S) instruments.
A wide range of data normalization strategies provide effectively equivalent performance in terms of SNR.

In Sec.~\ref{sec:total_uncertainty}, we present the results of a full pump-probe UED(S) dataset taken on polycrystalline monoclinic \votwo under conditions requiring the $P_0$ method to correct for lost counts. We demonstrate the impact of $P_0$ counting on the measured signals and validate the quantitative model for experimental uncertainties developed in Sec.~\ref{sec:pixel_response} in real pump-probe data.  
We conclude by suggesting ways that HPCD technology could be improved for ultrafast applications. 

\section{Methods} \label{sec:methods}
The key features of the experimental setup used to determine Dectris Quadro performance in UED(S) applications are described below, focusing on the properties of the electron source and detector.

\subsection{Hybrid Pixel Counting Detector} \label{sec:HPCD}
All hybrid pixel counting detectors~(HPCD) in use have two main components, which are bump bonded together: (i)~a high-resistivity semiconductor layer with a pixelated pn-diode structure and (ii)~a readout ASIC~(application-specific integrated circuit) array\cite{dectrismanual}.
The Dectris Quadro has $512\!\times\!512$ pixels, each $75\!\times\!75$\,\textmu m$^2$.
Incident electrons (or x-rays) deposit energy into a pixel in the sensor layer, creating electron-hole pairs.
An applied bias voltage across the layer collects those charges and results in a small current.
Every pixel has its own amplification, discrimination, and counting circuit, dramatically improving readout noise and dark current in recorded images\cite{foerster2019}.
An electron count is registered when the signal level generated by a pixel rises above a threshold value.
This threshold can be set at different levels in software depending on the incident electron energy.
A fundamental drawback of the counting electronics is the dead time following the registration of a count (100\,ns for the Dectris Quadro), during which additional counts cannot be registered.
Thus, 1/dead time sets the nominal maximum count rate that can be registered per pixel in CW beam applications (1107\,EPP).
However, the Quadro detector (and other HPCD detectors) features a so-called instant retrigger mode designed to improve high-rate counting by measuring the time above-threshold for the pixel signal, which correlates with the number of electrons hitting the pixel inside the dead time window\cite{Loeliger2012The}.
This provides a means to reduce count losses from pulse pileup in CW applications.
Here we test the retrigger mode for ultrafast pulsed exposures and demonstrate that it is ineffective at reducing pulse pileup-related count losses under these conditions and introduces noise to the counting.

\subsection{Electron Source}
The radio frequency compressed ultrafast electron scattering instrument used in these studies operates at 90\,keV electron beam energy and is described in further detail elsewhere\cite{Chatelain2012Ultrafast, otto2017solving}.
Electron pulses containing anywhere from 0 to $10^7$ electrons can be generated up to a maximum pulse repetition rate of 1\,kHz (laser-limited) via photoexcitation of a solid copper photocathode with 260\,nm UV pulses (4.77\,eV photon energy).
The UV pulses are generated by third harmonic generation from a 1\,kHz Ti:Sapphire chirped pulse amplified laser seeded by a frequency-doubled erbium fiber laser.
At the detector, these electron pulses are $<10$\,ps in duration, approximately four orders of magnitude shorter than the detector dead time, resulting in an effectively instantaneous exposure from the perspective of the readout electronics.
Ultrafast exposure conditions represent the worst-case scenario from the perspective of pulse pileup when multiple electrons hit a pixel inside a single shot.
The time between pulses (1\,ms) by contrast is four orders of magnitude longer than the dead time, which means that pulse pileup is not a concern between shots up to repetition rates of ~1\,MHz.

The Quadro has a triggerable maximum full-frame 16-bit readout rate of 2250\,Hz\cite{dectrismanual} that is more than sufficient to acquire the images produced by single electron pulse exposures at the maximum repetition rate of our experiments.
Single-shot images are essential for testing the counting behavior of HPCD pixels in all readout modes during ultrafast exposures over a range of electron dose conditions.

\section{Results}
In this section we present results on the performance of HPCDs subject to ultrafast exposures over a range of electron dose conditions when operating in normal and retrigger modes.
In Sec.~\ref{sec:pixel_response}, we present data on the counting behavior of individual pixels over a wide range of EPP conditions and establish an unconventional estimator (the $P_0$ counting method) that corrects for lost counts and extends the dynamic range of the detector at the high-end (saturation) by more than an order of magnitude.
In Sec.~\ref{sec:source_noise}, we present the full measured experimental noise spectrum of the UED(S) instrument employed and quantitatively compare data normalization strategies down to shot-to-shot normalization.
Finally, in Sec.~\ref{sec:total_uncertainty}, we present results of a full pump-probe UED(S) experiment using a HPCD and the $P_0$ counting method that quantitatively validates the model for experimental uncertainties developed.

\begin{figure}[t]
\includegraphics[width=1\columnwidth]{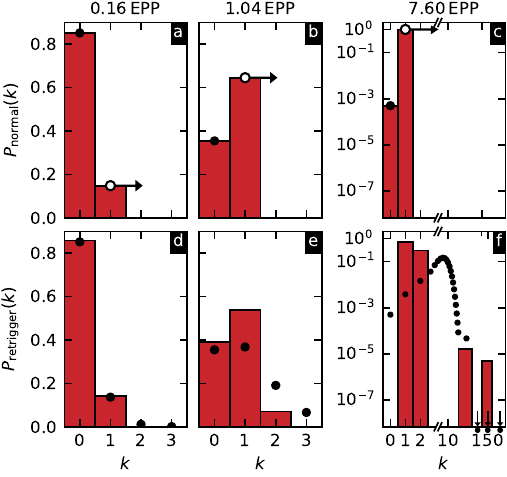}
\caption{Histograms of normalized single-pixel count distributions $P_\mathrm{normal}(k)$ in normal (\textbf{a--c}) and $P_\mathrm{retrigger}(k)$ in retrigger (\textbf{d--f}) counting modes under low (\textbf{a,d}), intermediate (\textbf{b,e}), and high (\textbf{c,f}) EPP; 0.16, 1.04, and 7.6\,EPP, respectively. $P_\mathrm{normal}(k)$ denotes the normalized probability distribution such that $\sum_k P_\mathrm{normal}(k) = 1$. The black points show the (expected) Poisson distribution for a mean value $\lambda$ = [EPP]. The open circles in (\textbf{a--c}) sum all counts k$\geq1$. Note that in panels c and f, the horizontal axis is linear for $k \leq 3$ and logarithmic for larger values. The black arrows in panel f indicate that the data points lie well below the lower limit of the y-axis.}
\label{fig:counts_vs_current} 
\end{figure} 

\subsection{Pixel response to ultrafast exposure}\label{sec:pixel_response}
Single-shot ultrafast exposures of the defocused, reasonably homogeneous electron beam were obtained under conditions that exposed large areas of the detector (i.e., many thousands of pixels) to electron dose conditions in the $10^{-3}$ to 10 EPP range at 1\,kHz pulse repetition rate.
To acquire images exposed by a single electron pulse, the detector acquisition time was set to 250\,\textmu s with acquisitions hardware-triggered by the laser amplifier system.
Typical single-pixel count distributions obtained under low (0.16\,EPP), medium (1.04\,EPP), and high (7.6\,EPP) average count rates acquired in normal and retrigger mode are shown in Fig.~\ref{fig:counts_vs_current}.
At 0.16\,EPP, pixel-count distributions are found to be in quantitative agreement with the Poisson distribution for $k=0$ and $k=1$ in both modes (shown in Fig.~\ref{fig:counts_vs_current}\,a,d), but in disagreement for $k=2$ count events.
In the 2,000 single-shot frames processed, zero two-count events were registered in normal mode (as expected in this mode of operation) and only two were registered in retrigger mode.
Fifteen two-count frames were expected at this average count rate.
The $k=1$ histogram bin obtained using normal mode should be understood as corresponding to any $k \ge 1$ event, so we compare the $k=1$ events against the predictions of the Poisson distribution integrated over $k \ge 1$ in Figs.~\ref{fig:counts_vs_current}\,a--c.

At 1.04\,EPP, the pixel count distribution in retrigger mode is in quantitative disagreement with the Poisson distribution at all count levels, although some $k=2$ events are clearly registered.
The histogram obtained in the normal mode is still in quantitative agreement with Poisson when the $k=1$ bin is understood as described above.
At 7.6\,EPP the typical-pixel count distribution obtained in retrigger mode shows dramatic departures from a Poisson distribution.
No frames containing 3--10 counts (or zero counts) are obtained, with some events (apparently random) registered with unphysical values exceeding 100.
It seems that the high instantaneous count rates induce not only massive overcounting artifacts, but also paralyze the pixels in subsequent images in a way that stops them from ever reading zero counts.
The histogram obtained in normal mode is, however, still in quantitative agreement with Poisson even under electron dose conditions where most frames register a count.

The sum of the retriggered histogram is approximately 3\% lower than that of the expected Poisson distribution in the low count case and approximately 37\% lower in the medium count case.
In the high count regime, such comparison is not meaningful due to the retrigger mode induced erroneous counts.
We conclude that the retrigger mode of the detector is not suitable for ultrashort electron pulse exposures and UED(S) applications.
Count losses are not significantly reduced in retrigger mode when ultrafast exposures are used, and this mode introduces random counting artifacts that add significant noise to the signals measured at all electron-dose levels when using ultrafast exposures.
These artifacts can only be (partially) corrected by processing all single-shot images.
Thus, all remaining data presented in this article were acquired in the normal counting mode of the Dectris Quadro.
A more detailed analysis of the implications of using the retrigger mode is presented in Appendix~\ref{sec:detector_counting_details}.

We showed that only normal-mode readout of HPCDs provides artifact-free counting in ultrafast applications.
However, even in normal mode, only zero count events can be relied on quantitatively (i.e., pixels that register a count should be understood as having measured 1 or more counts).
Given these constraints, we discuss strategies to optimize measurement uncertainties when using a direct electron detector in ultrafast applications.
First, for a single pixel and then for scattering signals across detectors.
\subsubsection{Single Pixel Counting}\label{sec:single_pixel_counting}
\begin{figure*}[t]
\includegraphics[width=1\textwidth]{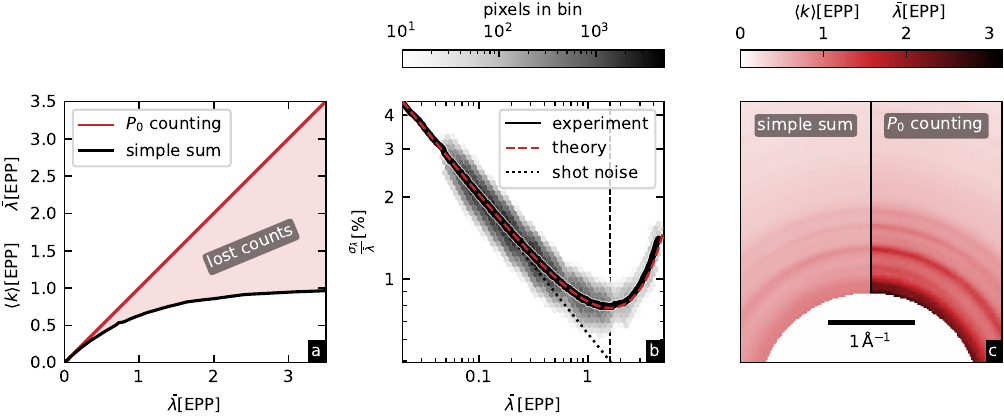}
\caption{Comparison of simple sum and $P_0$ counting. \textbf{a}~Detected per-pixel electron count rate in EPP using simple summation (black) and $P_0$ counting (red) vs the rate estimated by $P_0$ counting $\bar\lambda$ from single-shot diffraction patterns of polycrystalline monoclinic \votwo (described in text). \textbf{b}~Relative uncertainty in the per-pixel electron count rate vs $\bar\lambda$ for $P_0$ counting. The background heatmap shows the standard error in the mean obtained for every pixel in the dataset (logarithmic color bar indicated at top). A moving average of the experimental data is shown as a black solid line. The theoretical prediction from Eq.~\ref{eq:unc_in_px} is shown as a red dashed line. The theoretical optimum (minimum uncertainty) electron count rate is marked with a vertical dashed line. The contribution from the fundamental shot-noise limit decreases with increasing beam current and is plotted as a dotted black line. \textbf{c}~Diffraction pattern comparing simple sum and $P_0$ counting under the exposure conditions of the experiment, which yields $\bar \lambda > 0.5\,\mathrm{EPP}$ for many pixels. The left-half image was processed with pixel intensities determined by simple summation and subject to significant saturation due to lost counts. The right-half was processed by analyzing the ratio of zero events $\hat P_0$ and extracting $\bar\lambda$ to determine pixel intensities.}
\label{fig:counting}
\end{figure*}

In a single-shot, normal-mode image, all pixels have a value of either 0 or 1 even at high electron dose where Poisson statistics predict a high likelihood of multiple electrons arriving at the pixel in a single shot, as shown in Figs.~\ref{fig:counts_vs_current}\,a--c.
In this section, we address the question of the most reliable method for estimating pixel signal levels under such constraints and the limits to dynamic range with current HPCDs in ultrafast applications.

Since each pixel has its own readout ASIC, we can treat individual pixels as independent detectors and derive some fundamental properties.
The Poisson distribution determines the probability of detecting $k$ electrons in a pixel per shot when the mean value is $\lambda$ (i.e., $[\lambda]=\mathrm{EPP}$):
\begin{equation}
    P_k(\lambda) = \frac{\lambda^k \e^{-\lambda}}{k!}. \label{eq:poisson}
\end{equation}
With a binary detector that reports a value of 0 for $k=0$ and a value of 1 for all $k>0$, simple integration of detector counts leads to increasingly large losses of true electron counts for doses > 0.1\,EPP when multiple electron-count shots become common.
However, zero-count events are reliably measured by the detector well into this range as shown in Fig.~\ref{fig:counts_vs_current}\,a--c.
Thus, the fraction of zero-count events to total events provides a more reliable way to estimate the mean of the Poisson distribution for larger electron doses:
\begin{equation}
\lambda = -\ln(P_0)=-\ln\left(\frac{\#\mathrm{zeros}}{\#\mathrm{total}}\right) . \label{eq:estimator}
\end{equation}
In what follows, this will be referred to as $P_0$ counting for the determination of $\lambda$.
To estimate the uncertainty of the $P_0$ counting method, we define a Bernoulli variable 
\begin{equation}
    X_i = \begin{cases}
        1, & \text{if no electron detected} \\
        0, & \text{otherwise},
    \end{cases}
\end{equation}
and measure the fraction of zero events as the average over $N$ acquired pulses,
\begin{equation}
    \hat P_0=\frac{\#\mathrm{zeros}}{\#\mathrm{total}} = \frac{1}{N}\sum_{i=1}^N X_i.
\end{equation}
The variance of $\hat P_0$, a Bernoulli variable, is well known:
\begin{equation}
\Var(\hat P_0)= \frac{P_0(1-P_0)}{N}.
\end{equation}
Propagating the uncertainty on the estimator $\hat\lambda = -\ln(\hat P_0)$ yields the relative uncertainty
\begin{equation}
   \frac{\sigma_{\hat\lambda}}{\lambda} = \frac{1}{\lambda}\sqrt{\frac{1-\e^{-\lambda}}{\e^{-\lambda}N}}. \label{eq:unc_in_px}
\end{equation}
A minimum in the relative uncertainty is obtained at $\lambda_\mathrm{opt}=1.5936$, where
 $\frac{\sigma_{\hat\lambda}}{\lambda}(\lambda_\mathrm{opt}\!=\!1.5936, N) = \frac{1.2426}{\sqrt{N}}$.
This gives an effective upper bound on the dynamic range of current HPCDs in ultrafast applications.
Above an electron dose of 2\,EPP, the relative uncertainty rapidly increases with the $P_0$ counting method, diverging from the shot-noise limit, but remaining below 3\% up to 7\,EPP.
Below an electron dose of 1\,EPP, Eq.~\ref{eq:unc_in_px} asymptotes to the shot noise limit $\frac{\sigma_{\hat\lambda}}{\lambda} = \frac{1}{\sqrt{\lambda N}}$.

The predictions above for $P_0$ counting of $\bar\lambda$ are compared directly against measurements in Fig.~\ref{fig:counting}.
We obtained $N\!=\!25,\!200$ single-shot polycrystalline \votwo diffraction patterns at a repetition rate of 1\,kHz under electron-beam conditions that exposed the detector to EPP ranging from $\approx\!0.1$ to 3 at various positions in the pattern.
This range straddles $\lambda_\mathrm{opt}$, the expected minimum relative uncertainty region.
$\bar\lambda$ was determined for every pixel in the set of diffraction images using both simple sum and $P_0$ counting methods.
A comparison of those results is shown in Figs.~\ref{fig:counting} a and c.
At low $\bar\lambda$ both methods agree, but at higher $\bar\lambda$ they diverge due to pixel saturation and lost counts (simple summation scales as $1-\e^{-\bar\lambda}$).
The relative standard deviation of $\lambda$ was also determined for every pixel in the pattern.
These results, together with the predictions of Eq.~\ref{eq:unc_in_px} and shot noise, are shown in Fig.~\ref{fig:counting}\,b.
Agreement between these measurements and the predictions of Eq.~\ref{eq:unc_in_px} over the entire EPP range is excellent.
The determination of single pixel count rates and its associated uncertainty are accurately given by Eqs.~\ref{eq:estimator}~and~\ref{eq:unc_in_px}.

Comparing images produced by simple sum and $P_0$ counting in Fig.~\ref{fig:counting}\,c, it is evident that the diffraction ring contrast is visibly improved by $P_0$ counting, since this method corrects for the saturation and count losses present in bright pixels when employing a simple sum.
In Sec.~\ref{sec:total_uncertainty}, we demonstrate the associated impact on the resolvability of photo-induced changes to diffracted intensity for similar electron doses.

\subsubsection{Counting Across the Detector}
Polycrystalline materials or textured films generate diffraction patterns whose intensities only depend on the magnitude of the scattering vector $q=|\bm{q}|$.
This means that the intensity $I(q)$ is equivalently measured by multiple pixels located at different azimuthal angles with the same $q$.
By averaging over $M(q)$ physically equivalent pixels, the measurement uncertainty in the mean intensity improves by a factor of $\sqrt{M(q)}$ compared to the single pixel uncertainties shown in Fig.~\ref{fig:counting}\,b.
For the $\{(0\bar{1}1), (200), (0\bar{2}1), (220)\}$ diffraction rings of \votwo, our experimental geometry yields $\{574, 761, 872, 1106\}$  pixel-centers lying within a $\pm0.025\,\text{\AA}^{-1}$ band of each ring center.
Azimuthally integrating the diffraction pattern dramatically improves the measurement uncertainty in diffraction ring intensity.

\subsection{Additional Noise Contributions and Normalization}\label{sec:source_noise}
\begin{figure*}[t]
\includegraphics[width=1\textwidth]{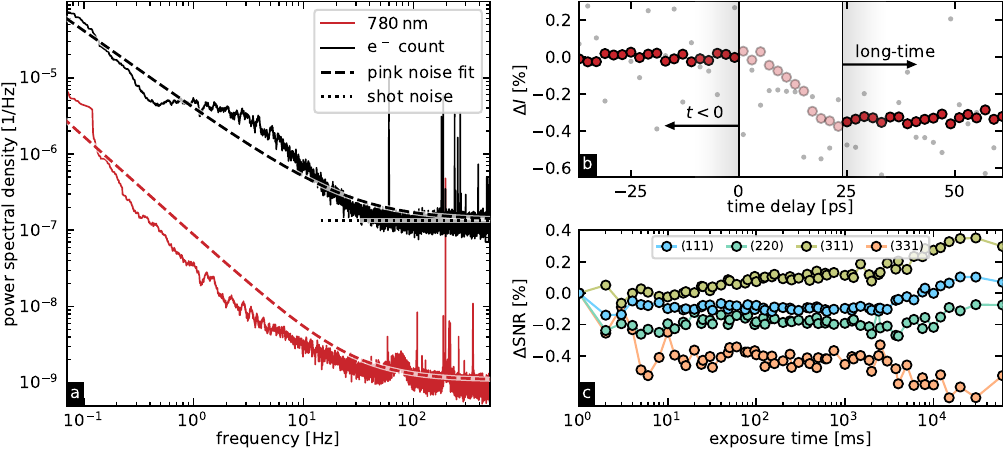}
\caption{\label{fig:normalization} Source noise and the impact of data normalization \textbf{a}~PSD of the (normalized) pulse-to-pulse variations in fundamental (780\,nm) laser amplifier output (red) and the total scattered electron count (black) in polycrystalline Pt diffraction patterns (acquisition conditions described in text). The black dotted line marks the shot-noise floor of the total electron-count signal. The black (red) dashed line shows the fit to an $1/f^\alpha$ noise model to the electron count (fundamental 780\,nm) data. \textbf{b}~Transient diffraction signal for the (220) Pt diffraction ring from $-30$ to $60$\,ps. For the SNR analysis, we use the time ranges outside the two black vertical lines. These data have been normalized with a shot-to-shot scheme. The gray points in the background show the same transient signal without using any normalization. \textbf{c}~Change in SNR vs simulated exposure time $\tau$ for four different Debye-Scherrer rings, with $\Delta \SNR = \left[\SNR(\tau)/\SNR(1\mathrm{ms})\right]\!-\!1$, indicates that SNR is effectively independent of the data normalization window over the range plotted. Note: The time resolution of the instrument (RF pulse compression) was not optimized for any of the measurements presented in this manuscript, since an unusually large range of electron bunch charge was being explored to test the detector. This has no impact on the detector results presented here, but the transient observed in panel b is strongly instrument broadened.}
\end{figure*}

A complete description of the SNR of electron scattering experiments requires a consideration of noise sources beyond the counting-related detection noise described in the previous section.
Kealhofer et al.\cite{kealhofer2019} investigated the signal-to-noise ratio of ultrafast electron diffraction experiments using a CCD detector and described the total noise in terms of a number of independent factors: 
\begin{equation}
\sigma_\mathrm{total} = \sqrt{\sigma_\mathrm{counting}^2 + \sigma_\mathrm{source}^2 + \sigma_\mathrm{gain}^2 + \sigma_\mathrm{int}^2 + \sigma_\mathrm{read}^2} \label{eq:total_noise}
\end{equation}

Some of the noise contributions that must be considered for a CCD detector are negligible for HPCDs, namely, gain, integration, and readout noise $\{\sigma_\mathrm{gain},\sigma_\mathrm{int},\sigma_\mathrm{read}\}$.
The per-pixel ASIC architecture of the HPCD reads images with virtually no readout and gain noise.
Noise due to integration is also not accumulated, since one image is recorded per electron bunch and the integration happens digitally.
Thus, only source and counting-related noise remain.
In Appendix~\ref{sec:detector_counting_details}, we provide a quantitative justification for this approximation of Eq.~\ref{eq:total_noise}.
Counting related noise is bound by shot-noise at low electron doses and by $P_0$ counting uncertainties at $\bar\lambda\!>\!1\,\mathrm{EPP}$ as demonstrated in the previous section (Fig.~\ref{fig:counting}\,b).
Here we investigate the relative contribution of source noise and counting noise to the total integrated electron scattering signal (undiffracted beam excluded) obtained from a 10\,nm-thick polycrystalline Pt film deposited on a 50\,nm thick SiN membrane TEM window $250\!\times\!250$\,\textmu m$^2$ in size (Norcada).
Simultaneously, we recorded the pulse energy of the laser-amplifier output with a photodiode and a boxcar integrator.

Over 500\,s, a total of $N = 500,\!000$ single-shot diffraction patterns and photodiode signals were collected.
Electron beam conditions were chosen such that the count rate within the analyzed region of the electron diffraction pattern (outside the main beam) ranges from 0.15 to 0.03\,EPP, well within the low electron-dose limit where count losses are negligible and uncertainties are determined by shot noise as shown in Fig.~\ref{fig:counting}.
The total diffraction pattern counts per shot (i.e., counts outside the transmitted beam) $\langle\mathcal{I}\rangle$ was 14,700 on average in this dataset.

The noise power spectral densities (PSDs) of both laser and electron signals $S(t)$ were obtained by taking the real Fourier-transform of the relative (normalized) intensity noise~(RIN) and are shown in Fig.~\ref{fig:normalization}\,a.
The overall PSDs can both be approximately modeled as pink noise, $S(f)=\frac{A}{f^\alpha} + C$, with an amplitude $A$ and an exponential scaling factor $\alpha$ and a constant offset $C$ (Fig.~\ref{fig:normalization}\,a dotted line).
The high-frequency noise floor of electron count signal is equal to the fundamental shot-noise limit, $\frac{2}{\Gamma \langle \mathcal{I}\rangle}$.
Using this value for the $C$-parameter, the best-fit values are $\alpha\!=\!1.02$ and $A\!=\!9\!\times\!10^{-8}$.
The (frequency-integrated) variance in electron counts is 1.1\%, of which shot noise accounts for $\approx\!60$\% of the variation and source noise $\approx\!40$\% at $\langle\mathcal{I}\rangle$ = 14,700.
As derived in Appendix~\ref{sec:electronsource_noise}, the variance of the total image counts can be written as
\begin{equation}
\Var(\mathcal{I}) = \bar{\mathcal{I}} + \epsilon^2\,\bar{\mathcal{I}}^2, \label{eq:source_noise_from_var}
\end{equation}
where $\epsilon$ represents common multiplicative bunch-charge fluctuations.
The first term is shot noise and the second is excess source noise, which grows quadratically with electron dose.
At our operating conditions this gives an effective relative intensity noise $\epsilon\!\approx\!0.65\%$.

The overall (integrated) variance in the fundamental laser signal is only 0.15\%.
By comparison, the excess electron source noise obtained from Eq.~\ref{eq:source_noise_from_var} is 0.65\%, approximately 4.3 times larger.
Given that the third-harmonic generation (converting NIR pulses into UV pulses) is expected to lead to a factor of 3 greater noise in the UV, there must be additional origins of source noise.
A plausible additional consideration are electron beam instabilities (mechanical and magnetic lens power supply instabilities) that lead to changes in the electron transmission through the sample window area.
We tentatively assign the additional source noise to these factors.

Normalization of acquired UED(S) images by the total number of electron counts is an important means to improve experimental SNR by reducing the impact of the electron source noise.
Here, normalization refers to self-normalization: dividing each image by its own total scattered intensity $\mathcal{I}$ (see the Appendix~\ref{sec:normalization_derivation} for a full quantitative development).
The ability to perform full-frame readout above 1\,kHz with HPCDs allows the exploration of shot-to-shot UED(S) data normalization strategies that have the potential to further improve SNR in UED(S) data, but have been impossible to implement with low frame-rate cameras due to relatively long required integration times and/or slow image readout.

Here, we comprehensively investigate the impact of the exposure time (number of pulses acquired per image) on the SNR of pump-probe UED(S) data to see if significant benefits are obtained by moving toward a shot-to-shot (self-)normalization scheme in these experiments.
We emphasize that the analysis in this section is entirely conducted in the low electron-dose limit, where count losses are negligible and simple summation of detector frames is a reliable measure of the electron dose.
The $P_0$ counting method is not required at such low fluences and is not feasible for low-exposure-time images.
To this end, we recorded a UED(S) dataset from a 10\,nm-thick polycrystalline Pt film on a freestanding 50\,nm SiN membrane, photoexcited with 150\,fs laser pulses centered at 780\,nm using a fluence of 0.1\,\mJcm at 1\,kHz repetition rate.
We recorded a total of $N\!=\!60,\!000$ single-shot diffraction images at 50 pump-probe time-delay points, allowing the direct comparison of exposure times from 1\,ms (single shot) to 1\,min (60,000 shots) using this single dataset.
Handling single-shot datasets of this size requires efficient storage and integration techniques; a suggested implementation approach used by the authors is outlined in Appendix~\ref{sec:data_management}.

From the transient differential diffraction intensity in a Debye-Scherrer ring
\begin{equation}
\Delta I(t) = \frac{I(t) - \langle I(t<0)\rangle}{\langle I(t<0)\rangle},
\end{equation}
we extract signal and noise to calculate the SNR per ring.
An example transient signal is shown in Fig.~\ref{fig:normalization}\,b for the (220) diffraction ring.
We note in passing that the exponential Debye-Waller response is slow and smeared out, because the time resolution of the instrument was not optimized as it is not of interest for the detector characterization conducted here.
Photoexcitation results in a rapid decrease in diffraction intensity due to the Debye-Waller effect, after which the diffracted intensity is approximately constant on the ps-timescale.
We quantify long-time signal as $\langle I(t>25\,\mathrm{ps})\rangle$ and the pre-photoexcitation intensity $\langle I(t<0)\rangle$ as a baseline.
The noise level is estimated from the standard deviation of the pre-photoexcitation time delays, yielding
\begin{equation}
\SNR = \frac{\big| \langle I(t>25\,\mathrm{ps})\rangle - \langle I(t<0)\rangle\big|}{\sqrt{\frac{1}{N_\mathrm{T}}\sum_{t<0} \left(I(t) -  \langle I(t<0)\rangle\right)^2}},
\end{equation}
where $N_\mathrm{T}$ is the number of pre-photoexcitation samples.

To simulate the exposure time dependence on SNR, a variable number of single electron shot diffraction patterns are first stacked/added before image normalization is performed and the transient differential intensity is computed.
Given the experimental conditions, we are limited to 60 different exposure time windows between 1\,ms and 1\,min, because the stack of 60,000 images must be divided into equal parts.
Overall results of this analysis are shown in Figure~\ref{fig:normalization}\,c.
We find that the SNR is constant to within 0.5\% over the entire (normalization) exposure-time range.
This shows that there is significant flexibility in the exposure time used in the acquisition of UED(S) data.
Shot-to-shot image normalization does not provide a significant SNR advantage in UED(S) experiments under typical source-noise conditions.
Shot-to-shot image acquisition/normalization followed by integration or integrated (multi-shot) image acquisition followed by normalization are similarly effective over a wide range of exposure time.

By contrast, simply eliminating image (self) normalization altogether reduces the SNR by up to a factor of 4 depending on the brightness of the diffraction ring considered.
Transient diffraction signals computed from unnormalized data are plotted on top of the normalized results in Fig.~\ref{fig:normalization} and the serious deterioration in SNR is clearly evident.
This work bridges the noise analysis performed in Kealhofer et al.\cite{kealhofer2019} and Claude et al.\cite{claude2025}, where the benefits of image normalization were clearly demonstrated, but only shot-to-shot\cite{claude2025} or long-exposure time windows\cite{kealhofer2019} were investigated.

This result is also in keeping with fundamental considerations under the assumption of (i) common multiplicative bunch-charge fluctuations, (ii) negligible additional cross-pixel correlations, and (iii) zero-mean jitter in a stationary process.
Appendix~\ref{sec:normalization_derivation} provides a derivation showing that the operation of normalization commutes with integration.

While the exposure time does not significantly impact the SNR acquired at low EPP, short integration windows can negatively impact the determination of absolute electron count rates with the $P_0$ method.
Because $P_0$ counting relies on $\hat\lambda=-\ln(\hat P_0)$, estimates of $\hat\lambda$ formed from too few frames are biased upward by the convexity of the logarithm; Appendix~\ref{sec:bias} quantifies this bias.
The practical implication is that $\hat\lambda$ estimates should be formed from a minimum of a few hundred shots, even though the normalization itself remains effective over the full exposure-time range.

In pump-probe measurements, however, we are primarily interested in changes in the scattering intensity.
When the two estimates use a matched number of events and have similar count rates, most of the logarithmic bias cancels in the difference signal. Since those conditions are met in all of our measurements, we will disregard the bias in the following.
Should these conditions not be met, the bias is still predictable and can be corrected analytically using the derivation presented in Appendix~\ref{sec:bias}.

\begin{figure}[t]
\includegraphics[width=1\columnwidth]{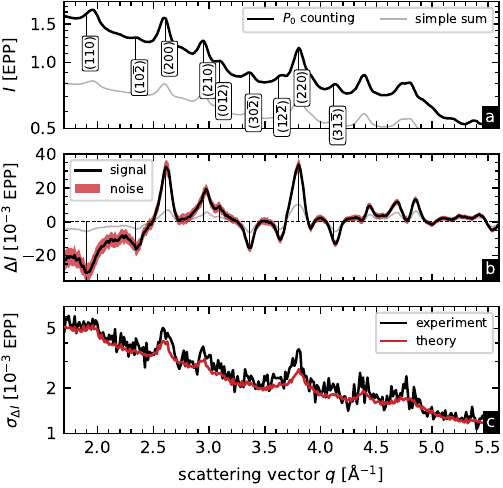}
\caption{Testing the $P_0$ counting method in a UED(S) experiment. \textbf{a}~Absolute azimuthally averaged diffraction pattern of \votwo in equilibrium. Selected diffraction rings are labeled with their respective Miller indices. \textbf{b}~Photo-induced signal change in diffracted intensity $>11$\,ps after photoexcitation (black) and rms noise (red). The light gray lines in the background of panels a and b show the signal acquired by simple summation, which are subject to count losses and saturation effects. \textbf{c}~Absolute uncertainty in the detected signal change as measured (black) and predicted by Eq.~\ref{eq:uncert_theo_full} (red).}
\label{fig:lineout_noise} 
\end{figure}

\subsection{Total Experimental Uncertainty} \label{sec:total_uncertainty}
Here, we present the results of UED(S) experiments conducted with a HPCD in the electron dose regime where $P_0$ counting becomes relevant in order to validate our model for the relative uncertainty in measured pixel intensities presented in Sec.~\ref{sec:single_pixel_counting}.
A UED(S) dataset was obtained for \votwo under 780\,nm photoexcitation (16.6\,\mJcm) and a (reduced) repetition rate of 200\,Hz.
The reduced repetition rate of 200\,Hz is required by the slower thermal recovery of \votwo, which cannot return to equilibrium within 1\,ms.
While the normalization analysis in Sec.~\ref{sec:source_noise} was framed in terms of integration time---the natural variable for a frequency-domain noise characterization---the quantity that governs the quality of the $P_0$ dose estimate is the number of acquired shots $N$ per stack.
At the reduced repetition rate of 200\,Hz used here, each stack of $N=500$ shots corresponds to an integration time of 2.5\,s, compared to 0.5\,s at 1\,kHz.
For each time-delay we recorded a stack of 500 single-shot diffraction images from which the electron dose for each pixel is determined by $P_0$ counting.
This acquisition was repeated seven times and all images are azimuthally integrated into 1000 $q$-bins as shown in Fig.~\ref{fig:lineout_noise}\,a.
Across all time delays, we average negative time-delay data $I(t<0)$ and long time-delay data $I(t>11\,\mathrm{ps})$, analogous to Fig.~\ref{fig:normalization}\,b.
The diffracted intensity before photoexcitation is shown in Fig.~\ref{fig:lineout_noise}\,a and the difference signal $\Delta I(q) = \langle I(t>11\,\mathrm{ps})\rangle-\langle I(t<0)\rangle$ in Fig.~\ref{fig:lineout_noise}\,b.
The applied normalization reduces the noise in the difference signal by a factor of 1.6 compared to unnormalized data.
We note that this improvement is smaller than in Fig.~\ref{fig:normalization}~b, because averaging over multiple time-delay points into two time bins already suppresses some of the source noise.

For the experimental uncertainty per $q$-bin, we compute the standard error of the difference of two independently averaged quantities.
Let $K_-$ and $K_+$ denote the number of stacks contributing to the pre- and long-excitation averages, respectively.
Furthermore, let $\sigma_\mp(q)$ be the standard deviations across stacks of the per-stack estimates in the $q$-bin.
The standard error of the difference is
\begin{equation}
\sigma_{\Delta I, \mathrm{exp}}(q) = \sqrt{ \frac{\sigma_-^2(q)}{K_-} + \frac{\sigma_+^2(q)}{K_+} }.\label{eq:difference_uncertainty}
\end{equation}
This experimentally measured uncertainty is shown as a red error band around the differential diffraction signal in Fig.~\ref{fig:lineout_noise}\,b.
To model the same quantity theoretically, we treat frame-to-frame bunch-charge fluctuations as Gaussian jitter of the instantaneous Poisson rate as discussed in Sec.~\ref{sec:source_noise}.
The key step is to combine the single-pixel uncertainty model with the source-noise assumptions made earlier.
In Appendix~\ref{sec:diffraction_uncert_w_source_noise}, we propagate the common multiplicative jitter from source noise through the $P_0$ estimator that is formed from $N$ events and averaged azimuthally over $M$ pixels.
We find that the uncertainty separates into a counting term that improves with the number of trials $NM$ and a source-noise term that remains correlated across pixels and hence only improves with $N$.
Applying the delta method yields the expected uncertainty for a $P_0$ counting based measurement in a given $q$-bin, over $K_\mp$ independent stacks gives (see Eqs.~\ref{eq:diffraction_feature_uncert_1} and \ref{eq:diffraction_feature_uncert_2}):
\begin{align}
\sigma_{\mp,\mathrm{th}} = \frac{\e^{\bar\lambda}}{\sqrt{K_\mp}}
\sqrt{\frac{\e^{-\bar\lambda + \frac{1}{2}\epsilon^2} - \e^{-2\bar\lambda + \epsilon^2}}{M\,N}
+\frac{\e^{-2\bar\lambda + 2\epsilon^2} - \e^{-2\bar\lambda + \epsilon^2}}{N}}, \label{eq:uncert_theo_full}
\end{align}
using the source-noise estimate $\epsilon = 0.65\%$ from Sec.~\ref{sec:source_noise} and using $M(q) \gg 1\:\forall\:q$.
In a last step, inserting Eq.~\ref{eq:uncert_theo_full} into Eq.~\ref{eq:difference_uncertainty} yields the full expression for the model uncertainty of a difference signal.
Panel~c overlays $\sigma_{\Delta I,\mathrm{exp}}(q)$ (black) with $\sigma_{\Delta I,\mathrm{th}}(q)$ (red).
The agreement across the entire $q$-range is good, confirming that the $P_0$ counting model accurately captures the statistical uncertainty and the data are shot-noise limited after normalization and aggregation over all frames of a stack, equivalent pixels, and time-delays.
Practically, this provides a simple planning tool for expected SNR in ultrafast electron diffraction experiments using HPCDs.

\section{Conclusions}
HPCDs can be a powerful tool for collecting UED(S) data without readout and dark noise, but HPCD pixels suffer saturation in ultrafast applications at the level of $\approx\!1$ detected electron per pulse.
The available retrigger modes on these detectors do not alleviate this problem and only introduce additional noise sources.
This places an important upper limit on the detector dynamic range, as counts/second in the brightest features cannot exceed the pulse repetition rate without introducing counting uncertainties in excess of shot-noise limits.
Ultrafast exposures subject HPCDs to profoundly different conditions than continuous beam experiments; i.e., the waiting time between subsequent counts in a pixel is exponentially distributed for continuous beams, but is effectively zero within a single ultrafast exposure.

Thus, these detectors excel when detecting very weak signals (phonon-diffuse) and collecting diffraction patterns with moderate dynamic range, such as gas-phase samples, polycrystalline materials, or textured films, where the scattered intensity is well distributed over the detector and even the brightest current-generation ultrafast electron sources can just reach a count rate of 1\,EPP.
Under such circumstances, detection of UED(S) signals at the shot-noise limit is easily obtained with HPCDs.
In single crystal experiments, however, Bragg peak features can easily saturate the detector and count losses adversely impact sensitivity.
Ideally, HPCDs with charge-integrating ASICs rather than threshold circuitry could be developed to reliably count multiple electrons inside a single shot, overcoming the limitations of this technology at the high end of the dynamic range for ultrafast applications.
Detector gating, which is straightforward in ultrafast applications through synchronization with the laser pulse train, could simplify such developments.

At source-noise levels typical of current generation UED(S) instruments, SNR in UED(S) is not significantly improved by shot-to-shot normalization enabled by the high frame rates.
SNR is virtually independent of the normalization time window.
As a practical consequence, once the detector response is characterized, single-shot acquisition is no longer necessary for routine UED(S) experiments.
The detector can integrate multiple exposures into a single readout frame without loss of data quality, dramatically reducing data storage requirements.

\section{Acknowledgments}
The authors acknowledge fruitful conversations with Mark Sutton (McGill University). L.~K. gratefully acknowledges support from a Fonds de Recherche du Quebec-Nature et Technologies (FRQNT) Merit fellowship.
B.~J.~S. gratefully acknowledges the support of the Canada Research Chairs program.
This work was supported by the FRQNT, the Canada Foundation for Innovation (CFI), and the National Research Council Canada (NRC) Collaborative R\&D program (Quantum Sensors).
L.~K., D.~C., M.~L. and B.~J.~S. benefit from their affiliation with the Regroupement québécois sur les matériaux de pointe (RQMP), see \url{https://doi.org/10.69777/309032}.

\section{Author Declarations}
\subsection{Conflict of Interest}
The authors have no conflicts to disclose.

\section{Data Availability}
The data that support the findings of this study are available from the corresponding authors upon reasonable request.

\bibliography{refs}

\appendix

\section{Electron Detector Counting Details\label{sec:detector_counting_details}}
This part of the appendix discusses the experimental details of electron counting, such as dark-count rates and unphysical pixel values.

\emph{Dark counts---}To quantify readout noise and dark counts in our setup, we take images without exposing the detector to electrons.
Over a total time span of 7\,h, 30 stacks of 15,000 images each were taken, resulting in 450,000 total images.
The total counts in those images are 852, detected randomly across the detector over time.
Spatially, one observes randomly distributed noise and a few streaks due to cosmic background radiation.
No pixel recorded more than two total counts.
On average, the probability of a pixel registering a dark count for a single exposure is
\begin{equation}
P_\mathrm{dark} = \frac{\#\mathrm{counts}}{\#\mathrm{images}\cdot\#\mathrm{pixels}}=7.6\times10^{-9},
\end{equation}
which is negligible compared to other sources of noise in the experiment.

%\label{sec:non-binary}
\emph{Non-binary events---}Approximately 1\% of the pixels in a dataset produce unphysical values of $2^{16}-1$ counts.
These defective pixels are masked during data processing.
Among the remaining pixels, only a very small fraction exhibit non-binary counts.
Specifically, we observe a fraction of $\frac{\#(N > 1)}{\#N}\!\approx\!5\!\times\!10^{-7}$.
During data processing, all such non-binary events that are not masked are set to a value of 1.

\emph{Retrigger mode---}The behavior of the retrigger mode of the electron detector has been discussed on a single-pixel basis in Sec.~\ref{sec:pixel_response}.
Especially noting the pathological values obtained from the detector in retrigger mode, we look at the behavior across the detector with respect to incident electron flux.
In the discussion we will use the count rate estimated with the $P_0$ counting method in normal mode as a baseline and obtain a reading in retrigger mode via a simple sum of all events.
The gain/loss can then be evaluated as $\eta = \frac{\bar\lambda_\mathrm{retr}}{\bar\lambda_\mathrm{normal}}-1$.
From the discussion in the main text, we expect both modes to behave similarly in the low electron flux regime; with increasing flux we expect an increase in lost counts, i.e., a decrease in $\eta$, since the electron bunches are much shorter than the pixel dead time.

\begin{figure}[t]
\includegraphics[width=1\columnwidth]{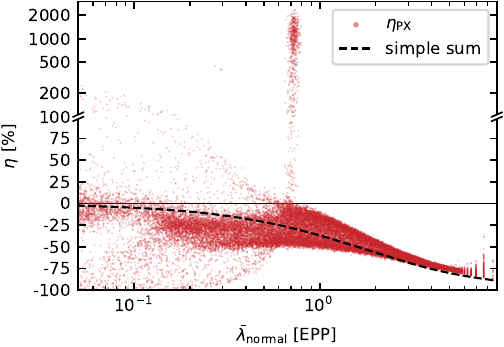}
\caption{Scatter plot for the gain/loss behavior when using retrigger vs normal trigger mode. Each scattered point corresponds to a pixel and the vertical axis is linear between $-100$ and 100\%, for larger positive values the axis is logarithmically scaled.}
\label{fig:retrigger}
\end{figure}

To investigate this, the gain/loss $\eta$ vs $\bar\lambda_\mathrm{normal}$ is plotted for every pixel in Fig.~\ref{fig:retrigger} using the same experimental conditions as in Sec.~\ref{sec:pixel_response}.
While an accumulation of points around $\eta=0$ can be seen, and a pronounced tail toward higher electron fluxes describes the loss of counts, there are many outliers.
Very pronounced is an almost vertical tail at 0.7\,EPP that shows pathological gains of up to 2,000\%.
We infer that those come from the pathological count that we presented in Fig.~\ref{fig:counts_vs_current}\,f of the main text.
As a guide the loss for a simple summation of counts in a binary detector is shown with a dashed line, since that is the expected model outcome for a retriggered pixel that fails to detect multiple counts.
Although in the regime of 1--6\,EPP there are only a few outliers, the loss is not as pronounced as it would be if the retriggered detector did not pickup any multi-electron doses.
This small effect is noticeable but not useful for conducting experiments.
The conclusion from investigating the \emph{instant retrigger technology} is that it is not suitable for applications in ultrashort compressed electron bunch environments and leads to corrupted data.

\section{Data management}\label{sec:data_management}
Collecting diffraction images at a 1\,kHz repetition rate produces large datasets, which require efficient storage and processing.
A stack of 60,000 images stored as unsigned 16-bit integers, as saved by the camera, has a size of approximately 31.5\,GB.
Even if each pixel is strictly binary, most software stacks treat a Boolean as the smallest addressable unit of memory---a full byte.
As a result, a Boolean array still uses eight times more memory than the information-theoretic minimum.
Because our detector frames are truly binary, we instead pack eight pixels in a single byte (bit-packing)\cite{c_book}.
This transformation is lossless and reduces the memory footprint of the image stack to approximately 1.97\,GB.
Arithmetic operations on a packed representation are less straightforward, but the $P_0$ counting method requires only summation.
Concretely, each byte encodes eight successive pixels; we read the byte value as an unsigned integer and count the number of ones using a population count operation.
For example, the pixel values
\begin{equation}
(0,1,1,0,1,0,0,0) \mapsto \texttt{0b01101000}=(104)_{10}
\end{equation}
are saved as a byte with decimal value 104.
To sum the number of ones in those eight pixels, we apply the population count operator and obtain
\begin{equation}
\operatorname{popcount}\left((104)_{10}\right)=3.
\end{equation}
The $\operatorname{popcount}$ operation is a precomputed lookup table returning the number of set bits for all integers between 0 and 255.
This method does not introduce significant performance degradation.

For storage on disk, we use HDF5 with chunking and in addition to bit-packing, lightweight compression filters further reduce the file size.
We find that the best combination of read/write performance and file size is achieved with packed binary data in combination with a bitshuffle compression filter\cite{bitshuffle}.
The final compression ratio vs the raw image array is 17.3.
If a minimal final file size is critical, gzip compression\cite{gzip} yields an even higher compression ratio of 23.7, but the read/write performance is 15 times slower.

\section{Electron source noise}\label{sec:electronsource_noise}
To determine how the electron source-noise contribution can be expected to scale with $\langle\mathcal{I}\rangle$ and the acquisition time, we consider multiplicative source noise to model its behavior.
We represent multiplicative source noise as a frame-wise scaling factor $Y$ that multiplies the entire diffraction pattern in a given exposure.
To keep the unconditional mean diffraction image unchanged, we take $\E[Y]=1$ and define the RIN of the electron source on the relevant timescales as $\epsilon^2=\Var(Y)$.
Conditioned on $Y$, we assume that pixel counts are independent, with
\begin{equation}
    I_{ij}\,|\,Y \sim \mathrm{Poisson}(Y\,\bar{I}_{ij}).
\end{equation}
Here $\bar I_{ij}$ is the mean intensity pattern in pixel $(i,j)$ in the absence of source fluctuations.
Under these assumptions, we obtain the expectation value $\E[I_{ij}\,|\,Y] = Y\bar I_{ij}$, variance $\Var(I_{ij}\,|\,Y)=Y\bar{I}_{ij}$, and covariance $\Cov(I_{ij},I_{uv}\,|\,Y)=0$ for $(i,j)\neq(u,v)$.
Thus, the total number of electron counts in an image is $\mathcal{I}=\sum_{i,j}I_{ij}$.
Using the law of total variance and covariance, we write
\begin{align}
\Var(\mathcal{I}) &= \sum_{i,j}\Var(I_{ij})+\sum_{(i,j)\neq (u,v)}\Cov(I_{ij},I_{uv}) \\
\intertext{and expand each term via conditioning on $Y$:}
\begin{split}
\Var(\mathcal{I}) &= \sum_{i,j} \left( \E_Y[\Var(I_{ij}|Y)] + \Var(\E_Y[I_{ij}|Y])\right) \\
&\quad + \sum_{(i,j)\neq (u,v)} \biggl( \E_Y[\Cov(I_{ij}, I_{uv} | Y)] \\
&\quad + \Cov(\E_Y[I_{ij}|Y], \E_Y[I_{uv}|Y])\biggr).
\end{split}
\intertext{The cross-pixel covariance $\E_Y[\Cov(I_{ij}, I_{uv} | Y)]$ vanishes after conditioning on the common jitter $Y$ and we simplify to}
\begin{split}
\Var(\mathcal{I}) &=\sum_{i,j}\left( \bar I_{ij} + \Var(Y\bar I_{ij}) \right) \\
&\quad+\sum_{(i,j)\neq (u,v)}\left( 0 + \Cov(Y \bar I_{ij}, Y \bar I_{uv}) \right).
\end{split}
\intertext{From here, we can group a shot noise term, a source-noise term, and a term from the cross-pixel covariance from source noise}
\Var(\mathcal{I}) &=\sum_{i,j} \bar I_{ij} + \Var(Y) \left(\sum_{i,j}\bar I_{ij}^2 + \sum_{(i,j)\neq (u,v)} \bar I_{ij} \bar I_{uv} \right) \\
&=\sum_{i,j}\bar I_{ij} + \sum_{(i,j),(u,v)} \bar I_{ij} \bar I_{uv} \epsilon^2 \\
\Var(\mathcal{I}) &= \bar{\mathcal{I}} + \Var(Y) \bar{\mathcal{I}}^2. \label{eq:shot-and-source}
\end{align}
Note that no assumptions were made about the frequency spectrum of the noise.
We assume only (i) multiplicative jitter and (ii) conditional Poisson counting.
Equation~\ref{eq:shot-and-source} separates shot noise that is linear in $\bar{\mathcal{I}}$ from excess source noise which is quadratic in $\bar{\mathcal{I}}$.

The source noise is a stationary fluctuating process, $Y(t)=1 + \delta(t)$, with PSD as measured (Fig.~\ref{fig:normalization}\,a and reasonably well approximated as:
\begin{equation}
S_\delta(f) = \frac{A}{f^\alpha} + C,\quad f\in [f_\mathrm{L},f_\mathrm{H}].
\end{equation}
The low-frequency cutoff $f_\mathrm{L}=\frac{1}{T}$ is the inverse of the acquisition time.
The high-frequency cutoff $f_\mathrm{H}=\Gamma/2$ is half the laser repetition rate.
The parameters $A$ and $\alpha$ are fitting parameters.
The constant baseline $C$ is fixed to represent the shot-noise floor for the electron counts.
The variance of the source noise can be calculated by integrating the PSD directly (Fig.~\ref{fig:normalization}\,a), or analytically via the model:
\begin{align}
\Var(Y) &= \Var(\delta) = \int_{\frac{1}{T}}^{\frac{\Gamma}{2}} \frac{A}{f^\alpha} + C \;\mathrm{d}f \\
 &= \frac{A}{1-\alpha} \left[ \left(\frac{\Gamma}{2}\right)^{1-\alpha} - \left(\frac{1}{T}\right)^{1-\alpha} \right] + C\left[\frac{\Gamma}{2} - \frac{1}{T}\right].
\end{align}

For true pink noise ($\alpha=1$) and $\Gamma \gg T^{-1}$ this simplifies to
\begin{equation}
\Var(Y) = \epsilon^2(T) = A\,\ln(f_\mathrm{H}T) + \frac{C\;\Gamma}{2},
\end{equation}

From the measurements shown in Fig.~\ref{fig:normalization}\,a we determine the laser source noise $\epsilon_\mathrm{L}\!=\!0.15\%$ and the electron source noise $\epsilon_\mathrm{e}\!=\!1.1\%$.

\section{Normalization w.r.t. Exposure Time}\label{sec:normalization_derivation}
In Sec.~\ref{sec:source_noise}, we showed experimentally that the normalization time-window does not affect the SNR of pump-probe signals.
Here, we provide a formal proof that, under multiplicative source noise and conditional Poisson counting, normalization commutes with integration.

Suppose we are conducting a measurement of a true diffraction signal $S_{ij}$ across the detector in pixel space $\{i,j\}$ with a total intensity $\sum_{i,j}S_{ij}=1$, probed with a mean electron bunch charge $\Qb$ and a shot-to-shot jitter $\delta_k$ in shot $k$.
The relative intensity jitter $R_\delta(m)$ can be related to the PSD of the source $\Phi_\delta(f)$:
\begin{align}
    R_\delta(m)&=\Cov(\delta_k,\delta_{k+m}), \\
    \Phi_\delta(f)&=\sum_{m=-\infty}^{\infty} R_\delta(m)\,\e^{-\i 2\pi f m / \Gamma},
\end{align}
with the integer shot distance $m$.
$\Phi_\delta(f)$ for our experiment is shown in Fig.~\ref{fig:normalization}\,a.
We assume no systematic drift in intensity $\langle \delta_k \rangle = 0$, and a stationary process---$R_\delta(m)$ is not a function of $k$---i.e., $R_\delta(m)$ only depends on the lag $m$.
Normalizing images over the intensity of an image stack of size $N$ results in a normalized image
\begin{equation}
    \Inorm_{ij}=\frac{T_{ij}}{T_\mathrm{tot}} = \frac{\sum_{k=1}^N I_{ijk}}{\sum_{k=1}^N \sum_{uv} I_{uvk}}.
\end{equation}
Under the assumption of small jitter $\delta_k \ll 1$, we apply the delta method and find that the pixel variance is
\begin{equation}
\Var(\Inorm_{ij})=\frac{1}{N^2\Qb^2}\bigg[ \Var(T_{ij}) + S_{ij}^2 \Var(T_\mathrm{tot}) - 2S_{ij} \Cov(T_{ij},T_\mathrm{tot}) \bigg],\label{eq:delta-method-normalization}
\end{equation}
neglecting any same-shot cross-pixel covariances beyond the multiplicative term---e.g., pixel readout crosstalk or charge diffusion between pixels---we formulate the cross-shot and within-shot (co)variances
\begin{align}
\Var(I_{ijk})           &=\Qb S_{ij} + \Qb^2S_{ij}^2R_\delta(0),\\
\Cov(I_{ijk}, I_{uvl}) &= \delta_{kl}\delta_{(i,j),(u,v)} \Qb S_{ij} + \Qb^2 S_{ij}S_{uv} R_\delta(k-l).
\end{align}
Treating variance and covariance terms in Eq.~\ref{eq:delta-method-normalization} separately, we find
\begin{align}
\Var(T_{ij}) &= \sum_{k=1}^N \Var(I_{ijk}) + \sum_{k\neq l}^N \Cov(I_{ijk}, I_{ijl}) \\
   &= N\Qb S_{ij} + \Qb^2S^2_{ij} \sum_{k,l=1}^N R_\delta(k-l). \label{eq:var_Tij}
\end{align}
\begin{widetext}
The variance term for the total intensity is
\begin{align}
\Var(T_\mathrm{tot}) &= \sum_{k=1}^N\Var\left(\sum_{i,j}I_{ijk}\right) + \sum_{k\neq l}^N\Cov\left(\sum_{i,j}I_{ijk}, \sum_{u,v}I_{uvl}\right)\\
&= \sum_{k=1}^N\left[\sum_{i,j}\Var(I_{ijk}) + \sum_{(i,j)\neq (u,v)} \Cov(I_{ijk}, I_{uvk})\right] + \sum_{k\neq l}^N\sum_{i,j}\sum_{u,v}\Cov(I_{ijk},I_{uvl}) \\
&= N\Qb + \sum_{k=1}^N\Qb^2 \sum_{i,j} S_{ij}^2R_\delta(0) + \sum_{k=1}^N \Qb^2R_\delta(0)\left(1-\sum_{i,j}S_{ij}^2\right) + \sum_{k\neq l}^N\Qb^2R_\delta(k-l) \\
&= N\Qb + \Qb^2 \sum_{k,l=1}^N R_\delta(k-l) \label{eq:var_Ttot}
\intertext{and for the covariance of the two we get}
    \Cov(T_{ij},T_\mathrm{tot}) &= \sum_{k,l=1}^N\Cov(I_{ijk}, \sum_{uv}I_{uvl})\\
    &= \sum_{k=1}^N \Cov(I_{ijk}, \sum_{uv}I_{uvk}) + \sum_{k\neq l}^N \Cov(I_{ijk}, \sum_{uv}I_{uvl}) \\
    &= \sum_{k=1}^N \left[\Var(I_{ijk}) + \sum_{(i,j)\neq(u,v)}\Cov(I_{ijk}, I_{uvk})\right] + \sum_{k\neq l}^N\sum_{u,v}\Cov(I_{ijk}, I_{uvl})\\
    &= \sum_{k=1}^N \left[ \Qb S_{ij} + \Qb^2S_{ij} R_\delta(0) \right] + \sum_{k\neq l}^N \Qb^2S_{ij}R_\delta(k-l) \\
    &= N\Qb S_{ij} + \Qb^2S_{ij} \sum_{k,l=1}^N R_\delta(k-l). \label{eq:cov_Tij_Ttot}
\end{align}
\end{widetext}
Plugging Eqs.~\ref{eq:var_Tij}, \ref{eq:var_Ttot} and \ref{eq:cov_Tij_Ttot} into Eq.~\ref{eq:delta-method-normalization} yields
\begin{equation}
\Var(\Inorm_{ij}) = \frac{S_{ij}(1-S_{ij})}{N\Qb}.
\end{equation}
All contributions from the power spectral density cancel out to first order.
The exposure time problem is commutative; the order of summation and normalization does not affect the noise performance.

To illustrate this, we choose $M$ to be a divisor of $N$ and form $K=\frac{N}{M}$ blocks of size $M$.
For each block $b=\{1,\dots,K\}$ one can (i)~integrate and then normalize
\begin{equation}
    T_{ij}^{(b)}=\sum_{k=(b-1)M+1}^{bM}I_{ijk},\quad
    I_{ij}^{\ordSN,(b)}=\frac{T_{ij}^{(b)}}{\sum_{u,v}T_{u,v}^{(b)}}
\end{equation}
or (ii)~normalize and then integrate
\begin{equation}
    I_{ij}^{\ordNS,(b)} = \sum_{k=(b-1)M+1}^{bM}\frac{I_{ijk}}{\sum_{u,v}I_{uvk}}
\end{equation}
to finally combine across blocks:
\begin{equation}
    \Inorm_{ij}=\sum_{b=1}^K I_{ij}^{\ordSN,(b)}
    \quad\text{or}\quad
    \Inorm_{ij}=\sum_{b=1}^K I_{ij}^{\ordNS,(b)}.
\end{equation}
Since each block contributes $\frac{S_{ij}(1-S_{ij})}{M\Qb}$ to the variance, and subdividing data into blocks does not introduce block-to-block correlations, the total variance is
\begin{equation}
    \Var(\Inorm_{ij}) = K\frac{S_{ij}(1-S_{ij})}{M\Qb} = \frac{S_{ij}(1-S_{ij})}{N\Qb}.
\end{equation}

\section{Bias of the 0/1 Electron Counter}\label{sec:bias}
The $P_0$ counting method introduced in Sec.~\ref{sec:single_pixel_counting} estimates the Poisson rate from the fraction of zero-count events via a nonlinear transformation.
Here we quantify the systematic bias this introduces when the number of frames $N$ used to form the estimate is small.

When repeating a 0/1 counting experiment $K$ times, the statistical uncertainty of the sample mean estimate $\bar\lambda$ improves with the number of samples taken $\frac{\sigma_{\bar{\lambda}}}{\bar{\lambda}} = \frac{1}{\sqrt{K}}\frac{\sigma_\lambda}{\lambda}$.
However, we also introduce a bias because of the nonlinearity of the logarithm.
By Jensen's inequality,
\begin{equation}
    \E[\hat\lambda] = \E[-\ln(\hat P_0)] \ge -\ln(\E[\hat P_0]) = -\ln(P_0)=\lambda.
\end{equation}
The convex nature of $-\ln(x)$ results in a systematic bias toward higher $\hat\lambda$.
An approximation of this bias can be derived using a second-order Taylor expansion of $f(x)=-\ln(x)$ around $x=P_0=\e^{-\lambda}$ and applying the shifting method\cite{ma2025}:
\begin{equation}
    \E[\hat\lambda] - \lambda = \frac{1}{2}\Var(\hat P_0)\ f^{\prime\prime}(P_0) = \frac{1-\e^{-\lambda}}{2N\,\e^{-\lambda}}.
\end{equation}
A measurement will be significantly biased when the number of events in the counting experiment $N$ is low.
The relative error of this bias drops below $10^{-3}$ for $N\gtrsim 1230$ events at $\lambda_\mathrm{opt}$.

\section{Relative uncertainty in a diffraction feature in the presence of source noise}\label{sec:diffraction_uncert_w_source_noise}
In this section, we will derive the relative uncertainty in a diffraction feature in the presence of source noise.
Adding source noise increases the variance and thus the relative uncertainty.

We extend the single-pixel uncertainty derived in Eq.~\ref{eq:unc_in_px} of the main body of the manuscript by the multiplicative frame-to-frame source-noise fluctuations described in Sec.~\ref{sec:source_noise} and averaging signals belonging to the same diffraction feature over multiple pixels.

From the law of total variance we know that
\begin{align}
    \Var(P_0) &= \frac{1}{N^2}\Var\left(\sum_i^N P_{0,i}\right) \\
    &= \frac{1}{N^2} \sum_i^N \left( \E[\Var(P_{0,i} | \Lambda_i)] + \Var (\E[P_{0,i} | \Lambda_i]) \right),
\end{align}
with the Poisson rates $\Lambda_i$ modeled as Gaussian distributed random variables with $\E[\Lambda_i]=\bar\lambda$ and $\Var(\Lambda)=\sigma_\mathrm{s}^2$, and the moment-generating function $m_k=\E[\e^{-k\Lambda}]=\e^{-k\bar\lambda+\frac{1}{2}k^2\sigma_\mathrm{s}^2}$.
We find that the variance of the ratio of zero-count events is
\begin{align}
\Var(P_0) &= \frac{1}{N} \left(m_1 - m_2 + m_2 - m_1^2 \right) \label{eq:shot_noise_var_in_px} \\
\Var(P_0) &= \frac{1}{N}\left(\e^{-\bar\lambda + \frac12\sigma_\mathrm{s}^2} - \e^{-2\bar\lambda + \sigma_\mathrm{s}^2}\right).
\end{align}
The first term is a counting uncertainty term that decreases with the number of events and the second term is only dependent on the electron flux and the strength of the source noise.
With the delta method we can write the relative uncertainty of the estimator $\hat\lambda=-\ln(\hat P_0)$ as
\begin{equation}
\frac{\sigma_{\hat\lambda}}{\bar\lambda}= \frac{1}{\bar\lambda} \sqrt{\frac{ \Var(\hat P_0)}{\E[\hat P_0]^2}} = \frac{\e^{\bar\lambda}}{\bar\lambda}\sqrt{\frac{1}{N}\left(\e^{-\bar\lambda + \frac12\sigma_\mathrm{s}^2} - \e^{-2\bar\lambda + \sigma_\mathrm{s}^2}\right)}.
\end{equation}

To move to the uncertainty of a diffraction feature, we define the source noise in a pixel $(i,j)$ as a fraction of the total source noise
\begin{equation}
\sigma_{\mathrm{s},ij} = \frac{\bar \lambda_{ij}}{\sum_{u,v} \bar \lambda_{uv}} \sigma_\mathrm{s, tot}.
\end{equation}
Pixel counts follow a Poisson distribution with rate $Y\cdot \bar{\lambda}_{ij}$, i.e., $N_{ij}\sim\mathrm{Poisson}(Y\cdot \bar{\lambda}_{ij})$, where the random variable $Y\sim\mathcal{N}(1,\epsilon^2)$, common to all pixels, is a Gaussian-distributed multiplicative factor representing source noise.

For a diffraction feature over the pixel domain $\mathcal{F}$, across $M$ pixels in $N$ frames (thus $NM$ trials), we derive the mean incident rate from the fraction of zero-events in the feature
\begin{equation}
P_{0,\mathcal{F}} = \frac{1}{N}\sum_k^N \frac{1}{M}\sum_{i,j}^M X_{ijk}.
\end{equation}
The variance of this quantity is
\begin{equation}
\Var(P_{0,\mathcal{F}}) = \frac{1}{N^2} \sum_k^N \frac{1}{M^2}\Var\left(\sum_{i,j}^M X_{ijk}\right),
\end{equation}
because the different frames are independent and different pixels in the same frame are subject to the same source noise.
Further,
\begin{align}
\begin{split}
\Var\left(\sum_{i,j}^M X_{ijk}\right) &= \underbrace{\sum_{i,j}^M\Var(X_{ijk})}_{\approx\text{ Eq.~\ref{eq:shot_noise_var_in_px}}} \\
&\quad+2 \sum_{1\leq (i,j) < (u,v) \leq M} \Cov(X_{ijk}, X_{uvk}), \\
\end{split}
\end{align}
with the already derived shot-noise variance for a single pixel and a covariance term between pixels that we can rewrite as 
\begin{align}
\begin{split}
\Cov(X_{ijk},X_{uvk})&=\underbrace{\E[\Cov(X_{ijk}, X_{uvk} | \Lambda_k)]}_{=0} \\
&\quad+ \Cov(\E[X_{ijk}|\Lambda_k], \E[X_{uvk}|\Lambda_k])
\end{split}\\
\rule{0pt}{18pt}&= \Cov(\e^{-\Lambda_k}, \e^{-\Lambda_k}) \\
&=\Var(\e^{-\Lambda_k})=m_2 - m_1^2
\end{align}
using the law of total covariances.
Finally, we arrive at
\begin{widetext}
\begin{align}
\Var(P_{0,\mathcal{F}}) &= \frac{1}{N^2} \sum_k^N \left[ \frac{1}{M^2} \left( \sum_{i,j}^M (m_1 - m_1^2) \right) + M(M-1) (m_2 - m_1^2) \right] = \frac{m_1 - m_1^2}{MN} + \frac{M-1}{M}\frac{m_2-m_1^2}{N} \\
&= \frac{\e^{-\bar\lambda + \frac{1}{2}\sigma_\mathrm{s}^2} - \e^{-2\bar\lambda + \sigma_\mathrm{s}^2}}{MN} + \frac{M-1}{M}\frac{ \e^{-2\bar\lambda + 2\sigma_\mathrm{s}^2} - \e^{-2\bar\lambda + \sigma_\mathrm{s}^2}}{N}.
\end{align}
\end{widetext}
Given the feature $\mathcal{F}$ contains many pixels, one can approximate $M\!\approx\!M-1$ and simplify further to
\begin{equation}
\Var(P_{0,\mathcal{F}}) = \frac{\e^{-\bar\lambda + \frac{1}{2}\sigma_\mathrm{s}^2} - \e^{-2\bar\lambda + \sigma_\mathrm{s}^2}}{MN} + \frac{ \e^{-2\bar\lambda + 2\sigma_\mathrm{s}^2} - \e^{-2\bar\lambda + \sigma_\mathrm{s}^2}}{N}.\label{eq:diffraction_feature_uncert_1}
\end{equation}
The first term improves with the total number of Bernoulli trials $MN$, whereas the second term only improves with $N$, because the multiplicative jitter is common to all pixels within a given frame.
Using the delta method, we compute the relative uncertainty
\begin{align}
\frac{\sigma_\mathcal{F}}{\bar\lambda} = \frac{\e^{\bar\lambda}}{\bar \lambda} \cdot \sqrt{\Var(P_{0,\mathcal{F}})}.\label{eq:diffraction_feature_uncert_2}
\end{align}

\end{document}